\documentstyle[epsf,fleqn,12pt]{aipproc1}
\renewcommand{\baselinestretch}{0.9}
\def\mbi#1{\mbox{\boldmath$#1$}}

\def\la{\langle}
\def\ra{\rangle}
\def\beeq{\begin{equation}}
\def\eneq{\end{equation}}
\def\beeqa{\begin{eqnarray}}
\def\eneqa{\end{eqnarray}}
\def\theequation{\arabic{equation}}

\def\thesection{\arabic{section}.}
\def\thesubsection{\arabic{section}.\arabic{subsection}}
\def\thepage{}

\newcommand\epsfigure[1]{
\vspace{8pt}
\begin{center}
\leavevmode\epsfbox{#1}
\end{center}
}
\addtocounter{section}{-1}
\begin{document}

\begin{center}

\mbox{}

\mbox{}

{\Large {\bf Optical Excitations in Hexagonal\\
Nanonetwork Materials}}

\mbox{}

\mbox{}

{\large Kikuo Harigaya}

\mbox{}

\mbox{}

{\small {\sl Nanotechnology Research Institute
and Research Consortium for Synthetic Nano-Function Materials Project, 
AIST, Tsukuba 305-8568, Japan}}

\end{center}

\begin{abstract}
Optical excitations in hexagonal nanonetwork materials, for example,
Boron-Nitride (BN) sheets and nanotubes, are investigated theoretically.  
The bonding of BN systems is positively polarized at the B site, and 
is negatively polarized at the N site.  There is a permanent electric 
dipole moment along the BN bond, whose direction is from the B site 
to the N site.  When the exciton hopping integral is restricted 
to the nearest neighbors, the flat band of the exciton appears at 
the lowest energy.    The symmetry of this exciton
band is optically forbidden, indicating that the excitons relaxed
to this band will show quite long lifetime which will cause
strong luminescence properties.
\end{abstract}

\mbox{}

\begin{center}
{\large {\bf INTRODUCTION}}
\end{center}

\mbox{}

The hexagonal nanonetwork materials composed of atoms with ionic
characters, for example, Boron-Nitride (BN) sheets and nanotubes [1],
have been investigated intensively.  They are intrinsically
insulators with the energy gap of about 4 eV as the preceding band 
calculations have indicated [2,3].  The possible photogalvanic 
effects depending on the chiralities of BN nanotubes have been 
proposed by the model calculation [4].  Even though optical 
measurements on the BN systems have not been reported so much, 
it is quite interesting to predict condensed matter properties 
of the hexagonal nanonetwork materials.

In this paper, we investigate optical excitation properties in 
BN systems.  The bonding is positively polarized at the B site, 
and is negatively polarized at the N site.  There is a permanent 
electric dipole moment along the BN bond, whose direction is 
from the B site to the N site.  The presence of the dipole
moments will give rise to strong excitonic properties as illustrated
in Fig. 1.  The energy of the highest occupied atomic orbital of N 
is larger than that of B, and the energy of the lowest unoccupied 
orbital of B is smaller than that of N.  Low energy optical excitations
are the excitations of the electron-hole pairs between the 
higher occupied states of N and the lower unoccupied states
of B atoms.

\begin{figure}[t]
\vspace{0.2cm}
\epsfxsize=4cm 
\epsfigure{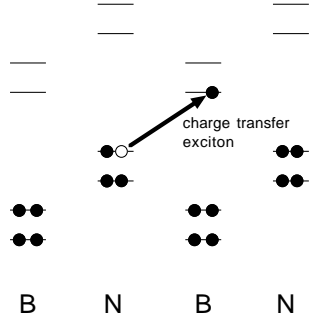}
\caption{%\footnotesize
Optical excitations along the BN alternations.}
\end{figure}%

\mbox{}

\begin{center}
{\large {\bf EXCITONS ON THE KAGOM\'{E} LATTICE}}
\end{center}

\mbox{}

We consider exciton interactions among nearest neighbor 
dipoles.  In Fig. 2 (a), the B and N atoms are represented by full and
open circles, respectively.  The several arrows show the directions
of dipole moments.  We assume one orbital Hubbard model with
the hopping integral of electrons $t$, the onsite repulsion $U$,
and the energy difference $\Delta$ between the B and N sites.  After
second order perturbations, we obtain the following forms of the 
nearest neighbor interactions: $J_1 = t^2 / (-\Delta + U)$ for the 
case of conserved excited spin (type-1 exciton) and 
$J_2 = t^2/\Delta + t^2/(-\Delta + U)$ for the case that spin of 
the excited electron flips (type-2 exciton).  The condition
$U > \Delta$ would be satisfied in general, and this means
that $J_1$ and $J_2$ are positive.  The interactions are present 
along the thin lines of Fig. 2 (a).  After the extraction of the 
interactions $J_1$ and $J_2$, there remains the two-dimensional 
Kagom\'{e} lattice which is shown in Fig. 2 (b).  Therefore, the 
optical excitation hamiltonian becomes:
\beeqa
H &=& \sum_{\la i,j \ra} \sum_{\sigma = \alpha, \beta} 
J_1 ( |i,\sigma \ra \la j,\sigma | + {\rm h.c.} ) \nonumber \\
&+& \sum_{\la i,j \ra}  
J_2 ( |i,\alpha \ra \la j,\beta | 
+ |i,\beta \ra \la j,\alpha | + {\rm h.c.} ), 
\eneqa
where the indices $i$ and $j$ mean the vertex points 
of the Kagom\'{e} lattice, and the sum is taken over the
nearest neighbor pairs $\la i,j \ra$ and the excited spin $\sigma$.  
The unit cell has three lattice points, 
namely, 1, 2, and 3, as shown in Fig. 2 (b).

The energy dispersions of the model are given in terms
of wavenumbers $\mbi{k} = (k_x,k_y)$:
\beeq
E =
\left\{ \begin{array}{l}
- 2 (J_1 + J_2), \\
(J_1 + J_2) [1 \pm \sqrt{1 + 4 \cos (k_x b /2)
[\cos (k_x b/2) + \cos (\sqrt{3}k_y b /2)]}],\\
2 (- J_1 + J_2), \\
(J_1 - J_2) [1 \pm \sqrt{1 + 4 \cos (k_x b /2)
[\cos (k_x b/2) + \cos (\sqrt{3}k_y b /2)]}],
\end{array} \right. 
\eneq
where the two dimensional x-y axes are defined as usual
in Fig. 2, and $b = \sqrt{3}a$ is the unit cell length
of the Kagom\'{e} lattice in Fig. 2 (b), and $a$ is the
bond length of Fig. 2 (a).  There appears a
dispersionless band with the lowest energy $-2 (J_1 + J_2)$.  
There is another dispersionless band at the higher energy
$2 (- J_1 + J_2)$.  The other four bands have dispersions 
which are like those of the two dimensional network of 
electrons on the graphite.  
Such the appearance of the flat band has been discussed
with the possibility of ferromagnetism in the literatures [5].
In the present case, the lowest optical excitation band
becomes flat in the honeycomb BN plane.  When the BN 
plane is rolled up into nanotubes, the flat band is 
dispersionless too.  Such the flat exciton band will have 
strong optical density originating from the huge density 
of states due to the weak dispersive character.

\begin{figure}[t]
\vspace{0.2cm}
\epsfxsize=5cm 
\epsfigure{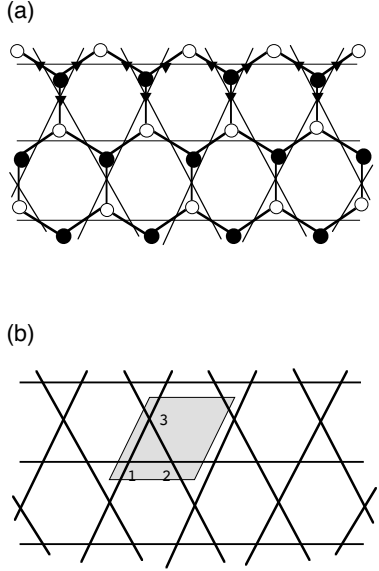}
\caption{%\footnotesize
(a) The hexagonal nanonetwork of boron (full circles)
and nitrogen sites (open circles).  Several arrows indicate
the directions of dipole moments, and the thin lines represent
the conjugate Kagom\'{e} lattice network. (b) The Kagom\'{e}
lattice extracted from Fig. (b).  The shaded area is the unit
cell, which have three lattice points indicated with numbers.}
\end{figure}%

\mbox{}

\begin{center}
{\large {\bf SYMMETRIES OF EXCITON WAVEFUNCTIONS}}
\end{center}

\mbox{}

\begin{figure}[t]
\vspace{0.2cm}
\epsfxsize=5cm 
\epsfigure{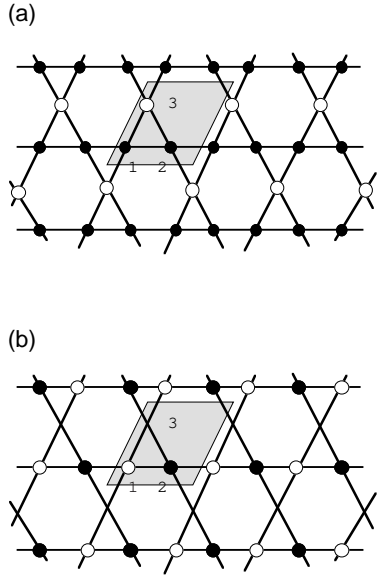}
\caption{%\footnotesize
Symmetries of two wavefunctions at $E=-2(J_1 + J_2)$.
The full and open circles indicate positive
and negative values at the lattice point, respectively.}
\end{figure}%

We look at symmetries of the lowest excitons with the
energy $- 2(J_1 + J_2)$.  The eigenvalue problem at the wavenumber 
$\mbi{k} = (0,0)$ gives twofold  degenerate solutions 
$\Psi_1^\dagger = (1/\sqrt{6}) (1,1,-2)$ and 
$\Psi_2^\dagger = (1/\sqrt{2}) (-1,1,0)$ both for 
the type-1 and type-2 exciton states.
The symmetry of the solution $\Psi_1$ is shown in
Fig. 3 (a), and that of the solution $\Psi_2$ is displayed in
Fig. 3 (b).  We find that both wavefunctions are symmetric
with respect to spatial inversion, and therefore they
have the symmetry {\sl gerade}.  The transition to the lowest 
exciton is optically forbidden.  The forbidden exciton state 
indicates that excitons relaxed to this lowest exciton band 
will show quite long lifetime which will cause strong luminescence 
properties.  In addition, the lowest energy excitons will
have huge density of state due to their flatness of the
band.  These properties might result in interesting
optical measurements in hexagonal nanonetwork materials.

\mbox{}

\begin{center}
{\large {\bf SUMMARY}}
\end{center}

\mbox{}

The flat band of the optically forbidden exciton appears 
at the lowest energy in the optical excitations of BN systems.  
The excitons relaxed to this band might show quite long lifetime 
which will cause strong luminescence properties.

\mbox{}

\begin{center}
{\large {\bf REFERENCES}}
\end{center}

\mbox{}

\noindent
{\footnotesize
1. D. Golberg, Y. Bando, K. Kurashima, and T. Sato, 
{\sl Solid State Commun.} {\bf 116}, 1 (2000).\\
2. A. Rubio, J. L. Corkill, and M. L. Cohen,
{\sl Phys. Rev. B} {\bf 49}, 5081 (1994).\\
3. X. Blase, A. Rubio, S. G. Louie, and M. L. Cohen,
{\sl Europhys. Lett.} {\bf 28}, 335 (1994).\\
4. P. Kr\'{a}l, E. J. Mele, and D. Tom\'{a}nek,
{\sl Phys. Rev. Lett.} {\bf 85}, 1512 (2000).\\
5. A. Mielke, {\sl J. Phys. A}, {\bf 24}, 3311 (1991); 
{\sl ibid.} {\bf 25}, 4335 (1992).\\
}

\end{document}